**Second-Order Vibrational Lineshapes from the Air/Water Interface**

Paul E. Ohno,[1] Hong-fei Wang,[2] Francesco Paesani,[3] James L. Skinner,[4] and

Franz M. Geiger[1]*

[1]Department of Chemistry, Northwestern University, Evanston, IL 60208, USA;

[2]Department of Chemistry and Shanghai Key Laboratory of Molecular Catalysis and

Innovative Materials, Fudan University, Shanghai 200433, China; [3]Departments of

Chemistry and Biochemistry and Materials Science and Engineering and San Diego

Supercomputer Center, University of California, San Diego, 9500 Gilman Drive, Urey

Hall 6218, La Jolla, CA 92093-0314, USA; [4]Institute for Molecular Engineering, Eck

hardt Research Center, Room 205, 5640 South Ellis Avenue, Chicago, IL 60637, USA.

*Corresponding author: geigerf@chem.northwestern.edu

**Abstract**. We explore by means of modeling how absorptive-dispersive mixing between the second- and third-order terms modify the imaginary $\chi^{(2)}_{total}$ responses from air/water interfaces under conditions of varying charge densities and ionic strength. To do so, we use published $Im(\chi^{(2)})$ and $\chi^{(3)}$ spectra of the neat air/water interface that were obtained either from computations or experiments. We find that the $\chi^{(2)}_{total}$ spectral lineshapes corresponding to experimentally measured spectra contain significant contributions from both interfacial $\chi^{(2)}$ and bulk $\chi^{(3)}$ terms at interfacial charge densities equivalent to less than 0.005% of a monolayer of water molecules, especially in the 3100 cm$^{-1}$ to 3300 cm$^{-1}$ frequency region. Additionally, the role of short-range static dipole potentials is examined under conditions mimicking brine. Our results indicate that surface potentials, if indeed present at the air/water interface, manifest themselves spectroscopically in the tightly bonded H-bond network observable in the 3200 cm$^{-1}$ frequency range.

**Introduction.** The air/water interface ranks among the most exhaustively probed systems in the field of nonlinear surface spectroscopy.[1-9] Recent reports of experimentally measured[10-13] and computed[14-17] vibrational sum frequency generation (SFG) spectra of the neat air/water interface show reasonably good matches, which has led to reports of new structural and dynamic insights into this fascinating yet enigmatic boundary of matter. Specifically, there is general agreement that the sharp positive feature in the Im($\chi^{(2)}$) spectra near 3700 cm$^{-1}$ represents dangling O-H oscillators pointing away from the water phase, while the broad negative feature from roughly 3200 to 3400 cm$^{-1}$ is due to H-bonded water with the O-H oscillators pointing into the water phase. However, significant controversy still remains regarding the existence and interpretation of a positive feature below 3100 cm$^{-1}$, as this feature is present in some theoretical[18-20] and experimental[10, 21] studies, while absent[13-15] or ambiguous[11, 16] in others.

Within the field of SFG spectroscopy, much recent attention has been devoted to understanding the relative contributions of $\chi^{(2)}$ and $\chi^{(3)}$ terms to computed and detected spectra. Such mixing has been shown to have important consequences for the spectral lineshapes observed in second-order vibrational responses from charged interfaces.[22-26] Here, we explore through modeling how $\chi^{(2)}$ and $\chi^{(3)}$ mixing modifies the imaginary $\chi^{(2)}_{\text{total}}$ responses from the air/water interface under conditions of varying minute charge densities, static dipole potentials, and ionic strength. In principle, when the input IR beam is p-polarized, the sign of the frequency-dependent amplitude of Im($\chi^{(2)}$) spectra informs on whether the z-component of the transition dipole moment in a given oscillator contributing to the SFG spectrum is directed net "up" or net "down".[27] Yet, as we will show, in the presence of interfacial charge, crossovers between positive and negative regions can occur, not caused by changes in orientation of interfacial molecules, but rather produced by $\chi^{(2)}/\chi^{(3)}$ mixing.

The mechanism of mixing between the potential-dependent contributions and the vibrationally resonant contributions involves three parts, as shown in the following expression for the measured SFG signal intensity, $I_{SFG}$:[26]

$$I_{SFG} \propto \left| \chi^{(2)}_{total} \right|^2 \tag{1a}$$

$$\chi^{(2)}_{total} = \chi^{(2)}_{NR} + \chi^{(2)}_{surf} + \frac{\kappa}{\sqrt{\kappa^2 + (\Delta k_z)^2}} e^{i\varphi} \Phi(0) \chi^{(3)} \tag{1b}$$

The first contribution is from the second-order non-resonant susceptibility, $\chi^{(2)}_{NR}$, generally taken to be real and containing no vibrational information as it is produced by the instantaneous electronic response of the system.[28-29] Recently, this source of nonlinear optical signal has been put to great use for self-heterodyning, or internally phase referencing, SFG and second harmonic generation responses.[24, 30] The second contribution, $\chi^{(2)}_{surf}$, is complex-valued, as it contains the vibrational responses of the surface species, each having vibrational transitions characterized by their Raman transition polarizabilities and IR transition dipole moments.[31-33] As the infrared frequency of the incident probe light sweeps across one of the vibrational resonances, this second contribution to the total SFG response increases to influence the SFG lineshape.

The two first contributions in eqn. (1) have formed the basis for much of the SFG lineshape analyses that have been published heretofore. Yet, it is now known[22-26] that the experimentally detected SFG response from charged interfaces contains an additional, third contribution stemming from a potential-dependent third-order response of the system under investigation. This third contribution has been shown to be of bulk origin.[24]

It is characterized by an effective third-order susceptibility, $\chi^{(3)}$, from any species that are polarized in the presence of an electrostatic potential, $\Phi(z)$, produced by any interfacial charges, and in principle contains *physical/chemical* contributions from molecules polarized in the presence of the static E-field as well as a purely *optical* third-order interaction between the static E-field and the incident light.[34] Though current efforts to understand the role of potential have focused on Coulombic potentials and mean-field theory, we caution that dipole potentials and other non-Coulombic potentials are of relevance as well. Given that the decay of the interfacial electrostatic potential $\Phi(z)$ with distance $z$ into the aqueous phase can extend tens or hundreds of nanometers away from the interface under conditions of low ionic strength, phase matching between the three waves needs to be taken into account, as described in Bloembergen and Pershan's early work on nonlinear optical signal generation from layered materials.[35] To do so, we apply the recently established formalism[22-26] that includes the inverse of the Debye screening length, $\kappa$, the inverse of the coherence length of the SFG process, $\Delta k_z$, and the interfacial potential, as shown in eqn. (1).

The $\chi^{(3)}$ phase angle, $\varphi$, does not refer to the phase of the various oscillators contributing to the resonant $\chi^{(3)}$ term, as those are defined separately in the sum over the oscillators constituting $\chi^{(3)}$ in the same manner as that discussed for $\chi^{(2)}$. Neither does $\varphi$ describe the phase of the "complex $\Psi$" discussed in reference (17), given that static potentials are real. Instead, the $\chi^{(3)}$ phase angle relates $\chi^{(2)}_{surf}$ and $\chi^{(3)}$, taking the form $\varphi = arctan\left(\Delta k_z / \kappa\right)$ for conditions where Gouy-Chapman theory applies (other expressions for how the surface potential may fall off with distance result in different solutions for the $\chi^{(3)}$ phase angle).[26] Moreover, when Gouy-Chapman theory applies, the $\chi^{(3)}$ phase angle $\varphi$ varies between 90° at low ionic strength and 0° at high ionic strength.[26] In the reflection geometries commonly used to probe the air/water interface, the importance of phase matching on the interfacial potential dependent term becomes largely negligible for ionic strengths exceeding roughly 1 mM, though purely absorptive mixing between $\chi^{(2)}$ and $\chi^{(3)}$ must still be considered.

Despite the care taken in experiments, ultrapure water is likely to contain adventitious organic carbon from, for instance, the occasional dust particle landing on a water surface, which may include surface-active species at concentrations below the ppb-level limit of detection of common Total Organic Carbon analyzers, such as the one accessible to us.[36] While SFG spectra of ultrapure water/air interfaces show no C-H oscillators from, say, alkyl tails of organic surfactants, their presence at small surface coverage levels is unlikely to be ruled out from those experiments, even when using heterodyne-detected SFG spectroscopy in the C-H stretching region.[37] Taken together, it is therefore worth considering what the consequences, if any, are of the presence of small amounts of interfacial charge (charged species equivalent to less than 0.005% of a monolayer of water molecules) on the SFG spectra produced at air/water interfaces. Calculating the number of water molecules per $m^2$ from the density of pure water at standard temperature and pressure, such a small percentage of ionization (0.005%) would correspond to an interfacial charge density of just 0.08 mC/$m^2$.

This estimate of adventitious charged surface-active species exceeds what would be expected for the limiting case of applying the auto-ionization of bulk water to the water surface at pH 7, and assuming, as a possible limiting case, that all ions of the same identity formed due to auto-ionization within a 2D sheet of water molecules reside within that sheet, while the balancing counter-ions would reside below the sheet. At pH 7, corresponding to 0.00000018% ionization, and using the bulk density of water at room temperature, there would be 1.8 x $10^6$ ions per $cm^2$, corresponding to 2.9 x $10^{-9}$ $C/m^2$. Local concentrations of ions at the air/water interface could of course be significantly higher or lower, provided a known surface propensity of hydroxide or hydronium. This consideration relates to whether the surface of water is acidic[38] or basic.[39-40]

We note that in a recent publication, Pezzotti $et$ $al$.[17] used 15-ps long DFT-MD simulations to directly calculate the surface $\chi^{(2)}$ and bulk $\chi^{(3)}$ contributions for a variety of aqueous interfaces. The lowest ionic strength considered was 0.4 M KCl, for which the Debye length (4.8 A) indicates that all $\chi^{(3)}$ contributions will be readily contained within the size of a typical simulation cell. However, for the low ionic strength and long Debye length present at the air/neat water interface, it is not practical to build simulation cells large enough to encompass the entirety of the diffuse layer. Thus, we present here a formalism that allows $\chi^{(2)}_{surf}$ spectra calculated from small simulation cells and in the absence of interfacial charge to be corrected to include the $\chi^{(3)}$ contribution, so that the computed spectra may be compared to the experimentally detected $\chi^{(2)}_{total}$ spectra. We also note that Pezzotti $et$ $al$. assumed that $\Delta k_z$ was 0, effectively setting the $\chi^{(3)}$ phase angle to 0. While this is a reasonable approximation for high ionic strength systems such as the ones they considered, experimental SHG[23-24] and SFG[22, 41] studies have shown that this phase interference effect must be taken into account to properly interpret SHG and SFG data at low ionic strengths. The formalism presented herein allows for the consideration of different $\Delta k_z$ values if needed for comparison with different experimental optical setups.

**Approach.** We follow our previously published method[26] for modeling spectral lineshapes produced from Equation 1 for a variety of conditions discussed in the Results section. However, unlike in that prior work, where we simulated the second- and third-order spectral contributions using Lorentzians, we use here a "lookup table" of published $\chi^{(2)}$ and $\chi^{(3)}$ spectra. Specifically, for our discussion, we use two computed $\text{Im}(\chi^{(2)}_{\text{surf}})$ spectra, namely the MB-pol (Fig. 1a) and E3B (Fig. 1b) derived spectra reported by the Paesani[14] and Skinner[15] groups, respectively. To account for the third-order contribution, we used the $\text{Re}(\chi^{(3)})$ and $\text{Im}(\chi^{(3)})$ spectra (Fig. 1c) published by Wen *et al.*[22] Given that the $\chi^{(3)}$ response primarily reports on molecules not localized directly at the interface and that that study showed negligible spectral variations with bulk pH or surface composition surveyed (fatty acid, long-chain alcohol), we assume here that it is appropriate to describe $\chi^{(3)}$ for the air/neat water interface as well. This assumption is also in agreement with the calculations of Pezzotti *et al.*, which show $\chi^{(3)}$ to be largely insensitive to the type of interface.[17] In order to cover the entire spectral range relevant for the OH stretches at the air/water interface (*i.e.* out to 3800 cm$^{-1}$), the $\text{Re}(\chi^{(3)})$ and $\text{Im}(\chi^{(3)})$ by Wen *et al.* were extrapolated linearly to zero for energies >3600 cm$^{-1}$ as a first-order approximation in the absence of published experimental data. This agrees reasonably well with the $\chi^{(3)}$ spectrum reported recently by Morita and coworkers;[42] regardless, we avoid interpreting spectral variations with potential we observe here in the free-OH region (around 3700 cm$^{-1}$) of the modeled spectra and instead focus on the lower frequency (H-bonded) OH stretching regions where the $\chi^{(3)}$ spectra has been experimentally measured. Due to the similarities between the calculated $\chi^{(3)}$ spectra of Pezzotti *et al*. and Morita and co-workers, and the experimental spectra we focus on here, qualitatively similar results are obtained in the analysis when the experimentally measured Im($\chi^{(3)}$) spectra are replaced with the calculated Im($\chi^{(3)}$) spectra. We also present in Fig. 1d the digitized experimental ssp-polarized Im($\chi_{total}^{(2)}$) spectra published by Sun *et al*.[11] and the one published by Nihonyanagi *et al*.[13] We note that while in overall good agreement with each other, the reported magnitudes of the Im($\chi_{total}^{(2)}$) amplitude in the low frequency region (3000 cm$^{-1}$ to 3200 cm$^{-1}$) and in the 3700 cm$^{-1}$ region seen in the experimental water surface spectra are not always consistent,[25] with the discrepancy in the low frequency region producing some controversy.

In order to quantitatively describe the interactions between the $\chi^{(2)}$ and $\chi^{(3)}$ terms, their relative magnitudes must be known. To do so, we normalize the hydrogen-bonded region of each spectrum to 1 and introduce a scaling factor, $s = \frac{Im(\chi^{(3)})_{\max}}{Im(\chi_{surf}^{(2)})_{\max}}$, as the ratio of the intensities of the un-normalized maxima in the hydrogen-bonded region. Reported literature intensities[11-13, 16-17, 22, 43] for Im($\chi^{(2)}$) and Im($\chi^{(3)}$) can be found in Tables S1 and S2, respectively. For what follows, we use the value calculated from the average of reported $\chi^{(2)}$ and $\chi^{(3)}$ values, namely s = 120 V$^{-1}$. We caution that it is currently not yet known whether the scaling factor may be frequency dependent, and to what extent such frequency dependence would be important. Moreover, it is not yet known what the role of laser field polarization plays in this problem, even though it may, given the recently reported differences in the salt concentration dependence of ssp- and pss-polarized SFG intensity spectra obtained from fused silica/water interfaces.[44]

For our analysis, we set the non-resonant $\chi^{(2)}$ response to zero for simplicity, i.e. we assume that its magnitude is negligible compared to the $\chi^{(2)}_{surf}$ term and the potential-induced $\chi^{(3)}$ term. We then add the appropriately scaled (*i.e.* multiplied by *s*) and phase shifted $\chi^{(3)}$ spectrum to the calculated $\chi^{(2)}_{surf}$ in order the model the $\chi^{(2)}_{total}$ spectrum that would be detected according to Eq. 1 to yield:

$$\overline{\chi^{(2)}_{total}} = \overline{\chi^{(2)}_{surf}} + s\,\overline{\chi^{(3)}}\,\frac{\kappa}{\sqrt{\kappa^2+(\Delta k_z)^2}}\,e^{i\varphi}\Phi(0) \qquad (2)$$

wherein the bars indicate normalized quantities. For both $\overline{\chi^{(2)}_{surf}}$ and $\overline{\chi^{(3)}}$, the normalization entails setting the maximum intensity in the hydrogen-bonded region equal to -1. The use of *s* therefore allows us to combine previously published data to explore how various conditions of large or small interfacial potentials and/or ionic strengths result in different detected $\chi^{(2)}_{total}$ spectra. As is common when reporting heterodyne-detected SFG spectra, we focus on Im($\chi^{(2)}_{total}$) due to its relevance for evaluating molecular orientations. Note that Equation 2 shows that the complex expression containing the $\chi^{(3)}$ phase angle, $e^{i\varphi}$, acts on both the real and the imaginary part of $\overline{\chi^{(3)}}$.

In what follows, we examine eqn. 2 by discussing its dependence on two important parameters in surface electrostatics, namely the surface charge density, σ, and the surface potential, $\Phi(0)$. Briefly, electrostatic potentials at interfaces can be described, for instance, by the familiar Gouy-Chapman model:

$$\Phi(0) = \frac{2k_BT}{ze}\sinh^{-1}\left[\frac{\sigma}{\sqrt{8k_BT\varepsilon_0\varepsilon_r n_i}}\right] \qquad (3)$$

The potential at the zero plane, $\Phi(0)$, depends on the thermal energy $k_BT$ and the valence z of the electrolyte, with *e* being the elementary charge, $\varepsilon_o$ being the vacuum permittivity, $\varepsilon_r$ being the relative permittivity of water, and $n_i$ being the concentration of ions. In our model, we choose $\varepsilon_r$ to be that of bulk water, even though other values may be relevant as well.[45] Eqn. 3 does not describe potentials relevant for charge-neutral surfaces, where dipole potentials can be important. In what follows, we also examine the role of dipole potentials. The annotated Mathematica notebooks available in the Supporting Information can be used to calculate the relevant parameters needed for the analysis presented next, or to simply input putative static dipole surface potentials if surface charge densities are not known.

**Results and Discussion.**

**Air/Neat Water Interface.** The $\chi^{(3)}$ phase angle for standard optical geometries can be estimated for the air/water interface using Gouy-Chapman theory. Laboratory studies of the air/water interface employ ultrapure deionized (DI) water, for which the autoionization of pure water puts a lower limit on the ionic strength (I=1 x $10^{-7}$ M). However, during the course of an experiment, atmospheric $CO_2$ may equilibrate with the aqueous phase and change the ionic strength and the pH in the bulk solution. Indeed, in a simple experiment (See Figure S1), we found following three and a half hours of contact with laboratory air the pH of pure (MilliQ system, resistivity = 18.2 MΩ) water to be 5.7 (I=2 μM). For a 2 μM 1:1 electrolyte concentration in water at 298 K, Gouy-Chapman theory predicts a Debye length (1/κ) of 210 nm. Assuming the commonly used incident angles of 45° for an 800 nm visible beam and 60° for the infrared beam and that the relative permittivity of the entire interfacial region optically probed, the extent of which is characterized by the ionic strength-dependent Debye length, remains that of bulk water (78), the $\chi^{(3)}$ phase angle is calculated to be near 78°. Note that the $\chi^{(3)}$ phase angle is invariant with interfacial charge density but depends instead on ionic strength (recall that

$$\varphi = arctan\left(\frac{\Delta k_z}{\kappa}\right).$$

For hypothetical interfacial charge densities corresponding to ±0.05 and ±0.01 mC/m$^2$, the surface potentials we calculate from Gouy-Chapman theory are ±15 and ±3 mV, respectively. These values are comparable in magnitude to what has been reported for the surface potential of the air/water interface.[46] We note that these values are the difference in potential between the bulk aqueous solution and the Gibbs-dividing surface, not the difference between bulk and vacuum, as has been reported to be several hundred mV.[16, 47-48] Fig. 2a-b shows results obtained for Im($\chi^{(2)}_{total}$) for $\chi^{(3)}$ phase angle, interfacial potential, and phase matching factor values corresponding to a total ionic strength of 2 μM and surface charge densities of 0 mC/m$^2$, ±0.01 mC/m$^2$, and ±0.05 C/m$^2$, modeled according to Equation 2, using the MB-pol derived (a) and E3B derived (b) surface spectra. Fig. 2 shows that at this ionic strength, the resulting lineshapes of the Im($\chi^{(2)}_{surf}$) spectra depend sensitively on the surface charge density. Moreover, the lineshapes change somewhat asymmetrically when changing the sign on the surface charge from negative to positive. Most strikingly, we find clear cross-overs from negative to positive values in the Im($\chi^{(2)}$) spectra occur between 3100 cm$^{-1}$ and 3300 cm$^{-1}$. Crossovers such as these and discrepancies among reported experimental and computed spectra have been the source of much controversy in the literature, as discussed above.

**Air/Brine Interface.** To examine the outcome of second- and third-order mixing in SFG responses from air/water interfaces under conditions of high ionic strength and surface (electrostatic or dipole) potentials, we recall the now classic adsorption studies of alkali ions at air/water interfaces[49-51] under brine conditions that show relatively large surface coverages of, for instance, iodide anions. Due to the increased polarizability of inorganic anions when compared to cations, these anions are generally considered to be surface active, whereas the smaller, harder cations are generally repelled from the interface, though this is a simplification of the often complex interactions that take place at the interface.[52-53] The electrostatic field, and associated potential, produced in the presence of a large number of anions at the air/water interface along with a layer of counterions located a few Å below can thus become significant. We also note that in addition to this Coulomb potential, dipole potentials associated with polarized water molecules oriented by the presence of the interface and associated electric fields would also become important under brine conditions (< 1 nm Debye lengths), as they act over just a short (<1 nm) distance when compared to the electrostatic (Coulomb) potential.

We thus modeled conditions of 1 M (Debye length = 0.3 nm) ionic strength and surface potentials of ±10 mV and ±5 mV, as shown in Figure 3. At this high ionic strength, the phase angle is approximately 0° and thus the mixing is nearly purely absorptive, yet substantial changes in the spectral lineshape can be seen even at the moderate potentials modeled here. Moreover, striking similarities are found between our results and those published by Allen and co-workers,[54-55] reinforcing their interpretation that observed changes in $Im(\chi^{(2)})$ spectra upon the addition of salt are largely driven by changes in the potential-dependent $\chi^{(3)}$ term. A full application of Equation 2 to data such as those published by Allen and co-workers opens up routes to quantitatively determining changes in interfacial potential, given that absolute spectral intensities are known.

**Conclusions.** In conclusion, we have used modeling to explore how absorptive-dispersive mixing between the second- and third-order terms contributing to second-order spectral lineshapes modify the imaginary $\chi^{(2)}_{total}$ responses from the air/water interface under conditions of varying minute Coulomb charge densities, σ, and sub-mM ionic strength and subsequent Coulomb potentials, Φ(0), that are on the order of several of mV. Additionally, short-acting dipolar surface potentials of similar values were examined for conditions of high salt concentrations as well to understand SFG lineshapes from interfaces containing stratified layers of alkali-halide ion pairs at air/brine interfaces. We used published $\chi^{(2)}$ spectra of the air/water interface that were obtained either from computations or from experiments, while published experimental $\chi^{(3)}$ responses were taken from a study of aqueous interfaces containing fatty acid monolayers, the sole currently available study reporting experimental $\chi^{(3)}$ spectra.

Our analysis shows that for the given set of experimental $\chi^{(2)}_{total}$ and $\chi^{(3)}$ spectra examined here, the Im($\chi^{(2)}_{surf}$) lineshapes are quite sensitive to absorptive-dispersive mixing for interfacial Coulomb charge densities lower than 0.005% of a monolayer of water molecules, with ionic strengths in the μM concentration regime. Clear cross-overs from negative to positive values in the Im($\chi^{(2)}$) spectra occur between 3100 cm$^{-1}$ and 3300 cm$^{-1}$ for these small charge densities. For brine conditions (1M salt), such cross-overs are produced for dipolar surface potentials as low as ±5 mV. Our analysis also indicates that dipole potentials, with their comparatively short distance dependence, are likely to be important under conditions of high (1M) ionic strength, for which the Debye length – a key determinant for the thickness of the SFG-active region – is on the order of a few Å. Yet, we caution that the results presented here indicate a need for reliably quantifying the $\chi^{(3)}$ scaling factor s and for determining whether it is subject to any frequency-dependence.

Furthermore, the vanishingly small amplitudes in reported $\text{Im}(\chi^{(2)}_{total})$ spectra between 3000 $\text{cm}^{-1}$ and 3300 $\text{cm}^{-1}$ would be consistent with the notion that Coulomb charge densities and surface dipole potential at the neat water/air interface are likely near or below 1 x $10^{-5}$ $C/m^2$, or that the $\chi^{(3)}$ scaling factor, s, is significantly overestimated in the reported literature values.

Our analysis effectively provides an upper limit on the possible charge density present at the air/neat water interface. This upper limit is three orders of magnitude larger than the estimation we provided earlier from the autoionization of water (2.9 x $10^{-9}$ $C/m^2$, *vide supra*). In the absence of any exogenous or adventitious carbon species, e.g. surfactants or contaminants, the upper limit of the charge density we provide here for the neat air/water interface would correspond to a net excess of ~6 x $10^9$ $H_3O^+$ or $OH^-$ ions per $cm^2$. In other words, $\text{Im}(\chi^{(2)}_{total})$ amplitudes that are statistically insignificantly different from zero in the 3000 $\text{cm}^{-1}$ to 3300 $\text{cm}^{-1}$ spectral region are consistent with the notion that the surface of water contains at most a net excess of either ~6 x $10^9$ $H_3O^+$ or ~6 x $10^9$ $OH^-$ ions per $cm^2$. Further improvements in the accuracy of the $\text{Im}(\chi^{(2)}_{total})$ amplitudes will help assess if the interfacial charge densities discussed here are in fact simply due to autoionization, as discussed in our upper limit estimation, and what role different surface propensities of hydronium or hydroxide ions may play to create a net acidic or net basic surface. However, what can be stated definitively is that $\text{Im}(\chi^{(2)}_{total})$ amplitudes in this region that are larger (*resp.* smaller) than zero are indicators of pronounced negative surface charge (*resp.* positive surface charge) at the air/neat water interface.

Improving the accuracy of the ratio $\frac{Im(\chi^{(3)})_{max}}{Im(\chi^{(2)}_{surf})_{max}}$, expressed here as the $\chi^{(3)}$ scaling factor s, opens the possibility to examine the low frequency region of the $Im(\chi^{(2)}_{total})$ spectra reported for air/water interfaces from the perspective of absorptive-dispersive mixing, in addition to phase dispersion and or the choice of the reference material.[12-13] Moreover,

our modeling study suggests it may be possible to spectroscopically measure the surface

potential of aqueous interfaces having magnitudes too small to be easily accessible with

other techniques, including Stark tuning rate measurements,[56] from the $Im(\chi^{(2)}_{total})$

amplitude in the 3200 cm$^{-1}$ frequency range, or at other heretofore uncharted frequencies,

provided the availability of reliable values for the scaling factor, s. Indeed, our results suggest that surface potentials manifest themselves spectroscopically in the tightly bonded H-bond network observable in the 3000 cm$^{-1}$ - 3200 cm$^{-1}$ frequency range. They may appear in other frequency ranges, such as the HOH bending frequency,[27, 57-59] as well.

**Data availability.** All relevant data are available from the authors upon request to the

corresponding authors, with the notebook used to generate the model spectra provided in

the Supplementary Information.

**Acknowledgments.** This work was supported by the US National Science Foundation

(NSF) under its graduate fellowship research program (GRFP) award to PEO. HFW

gratefully acknowledges support from the Shanghai Municipal Science and Technology

Commission (Project No. 16DZ2270100) and Fudan University. FP acknowledges

support from the NSF grant number CHE-1453204. This research used resources of the

Extreme Science and Engineering Discovery Environment (XSEDE), which is supported

by the National Science Foundation (Grant No. ACI-1053575). JLS acknowledges

support from the Institute for Molecular Engineering at the University of Chicago. FMG gratefully acknowledges support from the NSF through award number CHE-1464916 and a Friedrich Wilhelm Bessel Prize from the Alexander von Humboldt Foundation.

**Author Contributions.** PEO and FMG conceived of the idea. The data was analyzed and the manuscript was written with substantial contributions from all authors.

**Author Information.** The authors declare no competing financial interests. Correspondence should be addressed to FMG (geigerf@chem.northwestern.edu).

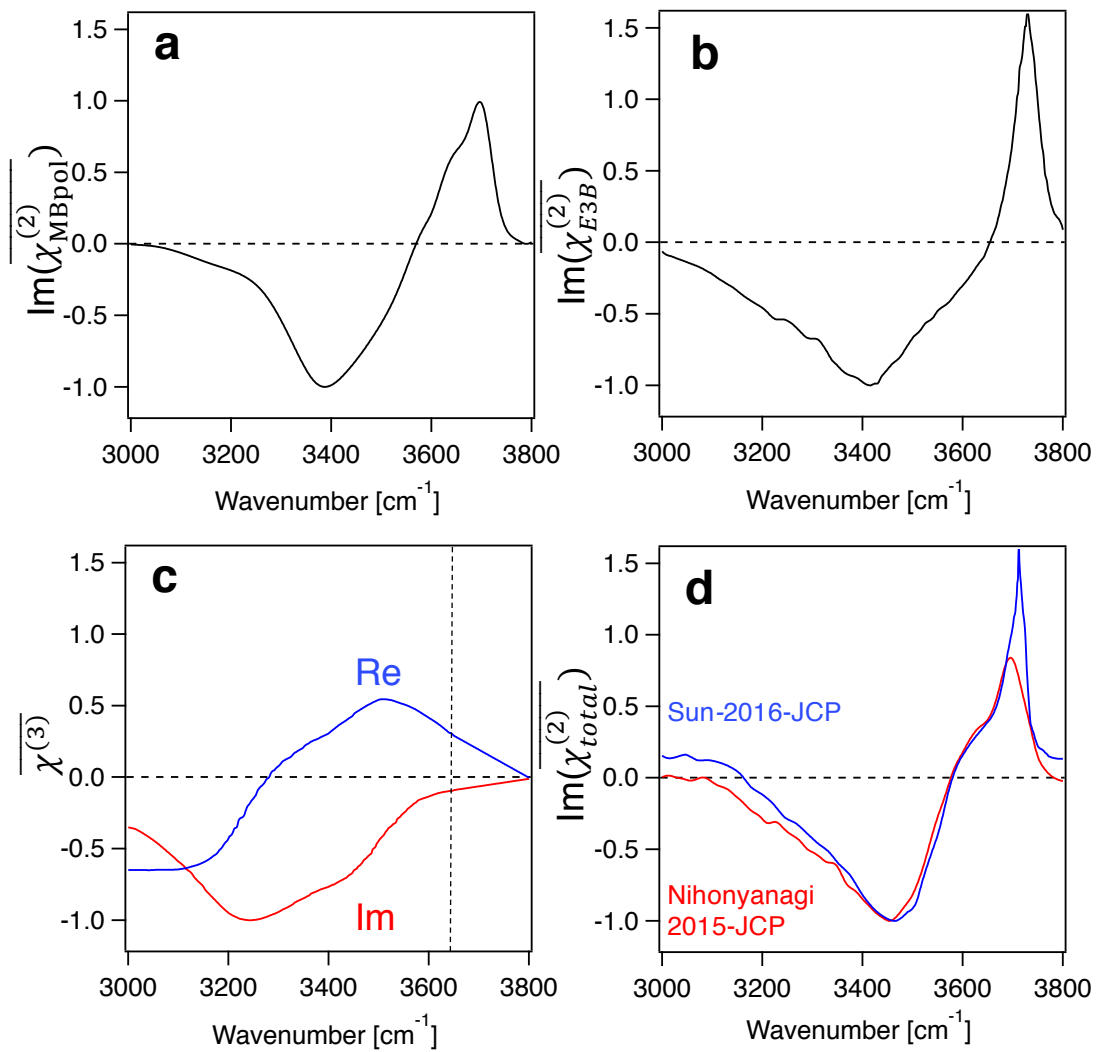

**Figure 1. Individual spectra used in the present model study.** (**a**) MB-pol derived Im($\chi^{(2)}$) spectrum, and (**b**) E3B-derived Im($\chi^{(2)}$) spectrum both normalized to their respective maximum intensities in the hydrogen-bonded region, Im$\overline{(\chi^{(2)}_{surf})}_{max}$ = -1. (**c**) Experimental Im($\chi^{(3)}$) and Re($\chi^{(3)}$) spectra reported by Shen, Tian, and coworkers in 2016,[22] normalized to Im$\overline{(\chi^{(3)})}_{max}$ = -1 at its maximum intensity in the hydrogen-bonded region and extrapolated linearly to zero from the vertical dashed line on. (**d**) Experimental ssp-polarized Im($\chi^{(2)}$) spectrum of the air/water reported by Tahara and coworkers in 2015 and Shen and co-workers in 2016, both normalized to their respective

maximum intensities in the hydrogen-bonded region, $Im(\overline{\chi^{(2)}_{total}})_{max} = -1$. Please see text for details.

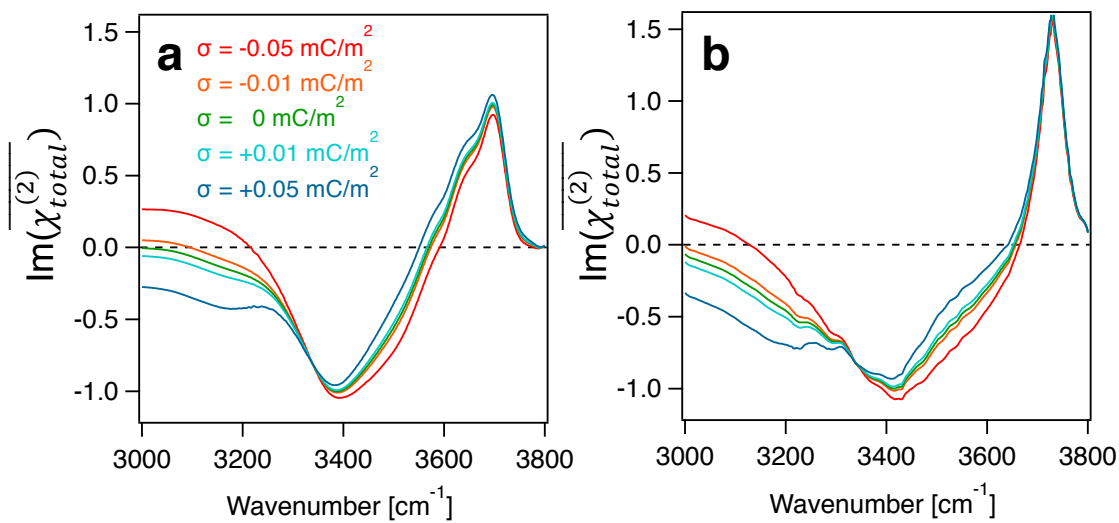

**Figure 2.** $\chi^{(2)}$ **and** $\chi^{(3)}$ **interactions for air/neat water interfaces.** $\mathrm{Im}(\overline{\chi^{(2)}_{total}})$ spectra calculated using Eqn. 2 for an ionic strength of 2 μM and charge densities of -0.05 mC/m$^2$ (red, corresponding to -15 mV and 0.003% of a monolayer charged), -0.01 mC/m$^2$ (orange, corresponding to -3 mV and 0.0006% of a monolayer charged), 0 mC/m$^2$ (green, corresponding to 0 mV), +0.01 mC/m$^2$ (light blue, corresponding to +3 mV and 0.0006% of a monolayer charged ), and +0.05 mC/m$^2$ (navy, corresponding to +15 mV and 0.003% of a monolayer charged) for (**a**) MB-pol and (**b**) E3B derived surface spectra. The scaling factor between the normalized surface and bulk spectra, s, was set equal to 120 V$^{-1}$, in agreement with reported literature values (see Supplementary Information Tables S1 and S2).

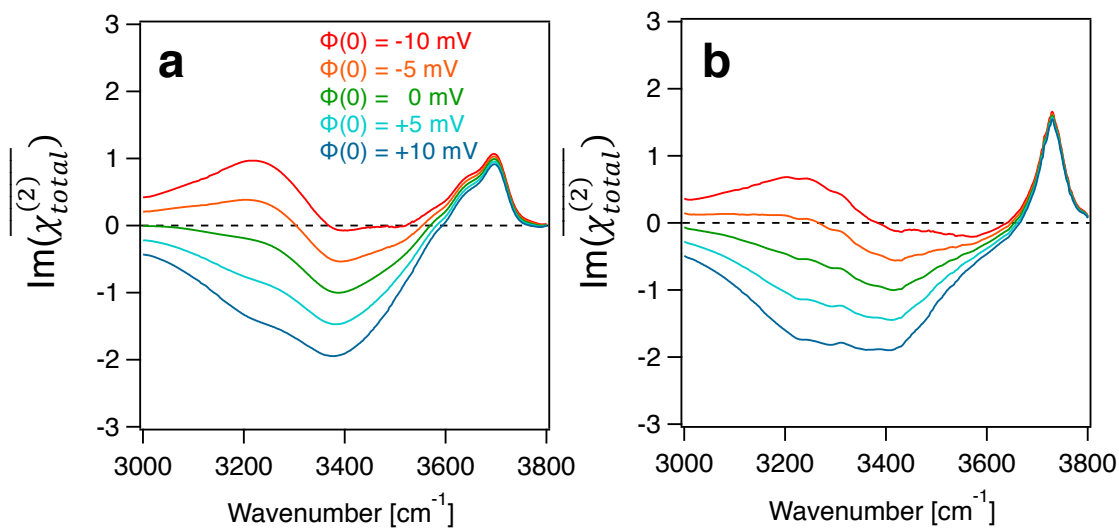

**Figure 3. χ$^{(2)}$ and χ$^{(3)}$ interactions for brine/water interfaces.** Im($\overline{\chi^{(2)}_{total}}$) spectra calculated using Eqn. 2 for an ionic strength of 1 M with surface potentials of -10 mV(red), -5 mV (orange), 0 mV (green), +5 mV (light blue), and +10 mV (navy) for (**a**) MB-pol and (**b**) E3B derived surface spectra. The scaling factor between the normalized surface and bulk spectra, s, was set equal to 120 V$^{-1}$, in agreement with reported literature values.